\documentclass[12pt]{iopart}

\usepackage{cite}
\usepackage{graphicx, subfigure}
\usepackage{latexsym}
\usepackage{color}

\begin{document}
\today

\title[Doppler tomography in fusion plasmas and astrophysics]{Doppler tomography in fusion plasmas and astrophysics}

\author{M~Salewski$^1$, B~Geiger$^2$, W~W~Heidbrink$^3$, A~S~Jacobsen$^1$, S~B~Korsholm$^1$, F~Leipold$^1$, J~Madsen$^1$, D~Moseev$^{2,4}$, S~K~Nielsen$^1$, J~Rasmussen$^1$, L~Stagner$^3$, D~Steeghs$^5$, M~Stejner$^1$, G~Tardini$^2$, M~Weiland$^2$ and the ASDEX Upgrade team$^2$}

\address{$^1$ Technical University of Denmark, Department of Physics, DK-2800 Kgs. Lyngby, Denmark\\
$^2$ Max Planck Institute for Plasma Physics, D-85748 Garching, Germany\\
$^3$ University of California, Irvine, California 92697, USA\\
$^4$ FOM Institute DIFFER, 3430 BE Nieuwegein, The Netherlands\\
$^5$ Department of Physics, University of Warwick, Coventry, CV4 7AL, UK}

\ead{msal@fysik.dtu.dk}

\begin{abstract}
Doppler tomography is a well-known method in astrophysics to image the accretion flow, often in the shape of thin discs, in compact binary stars. As accretion discs rotate, all emitted line radiation is Doppler-shifted. In fast-ion D$_\alpha$ (FIDA) spectroscopy measurements in magnetically confined plasma, the D$_\alpha$-photons are likewise Doppler-shifted ultimately due to gyration of the fast ions. In either case, spectra of Doppler-shifted line emission are sensitive to the velocity distribution of the emitters. Astrophysical Doppler tomography has lead to images 
of accretion discs of binaries revealing bright spots, spiral structures, and flow patterns. Fusion plasma Doppler tomography has lead to an image of the fast-ion velocity distribution function in the tokamak ASDEX Upgrade. This image matched numerical simulations very well. Here we discuss achievements of the Doppler tomography approach, its promise and limits, analogies and differences in astrophysical and fusion plasma Doppler tomography, and what can be learned by comparison of these applications.
\end{abstract}


\section{Introduction}
\label{sec:intro}
Doppler tomography has been used to image a fast-ion velocity distribution function in a fusion plasma \cite{Salewski2014}. While this application of Doppler tomography is in its infancy, it has been used to study astrophysical accretion discs for more than 25 years \cite{Marsh1988, Marsh2001, Marsh2005}. Readily observable accretion discs form in pairs of stars, called interacting binaries, in which matter flows from one star to its companion. Angular momentum tends to confine these discs within the orbital plane of the binary with the gas orbiting around the more massive, compact component in the system, often a stellar remnant. They form when this compact object pulls matter towards it. Angular momentum in the accretion disc is transported outwards, and hence matter spirals inwards and eventually reaches the accretor.  Astrophysical Doppler tomography has provided images of accretion discs for several classes of binaries \cite{Marsh2001, Marsh2005, Steeghs2004, Morales-Rueda2004, Schwope2004, Richards2004, VrtiLek2004}.

Magnetically confined laboratory plasmas are heated to $\sim 10$~keV mostly by fast ions generated by injected energetic neutrals ($\sim$ 30~keV - 1~MeV), by electromagnetic wave acceleration (up to MeVs), or finally in a fusion power plant by the DT fusion reaction (3.5~MeV). Fast ions are magnetically forced on twisted trajectories within the donut-shaped plasma until they become part of the thermal ions. At the tokamak ASDEX Upgrade we can generate a variety of fast-ion populations by using neutral beams or electromagnetic waves at the ion cyclotron frequency \cite{Gruber2007, Kallenbach2012, Stroth2013}.

Doppler tomography is analogous to standard tomography, but the images are constructed in velocity space rather than in position space. This is possible due to the Doppler shift of line radiation from emitters on trajectories with a near-circular component. Such trajectories are typical for rotating accretion discs of binary stars and gyrating ions in magnetized plasma that have locally helical trajectories. The velocity vector of the emitter and its projected velocity $v_{LOS}$ onto the line-of-sight (LOS) of a detector depend on the angular position of the emitter in its orbit with respect to the line-of-sight. We refer to this angle as phase angle $[0,2\pi]$ or alternatively as phase $[0,1]$ following the astrophysical literature. The wavelength shift of emitted photons is proportional to $v_{LOS}$ according to the Doppler shift $\lambda - \lambda_0 = v_{LOS} \lambda_0/c$ where $c$ is the speed of light, $\lambda_0$ is the unshifted wavelength of the line emission, and $\lambda$ is its Doppler-shifted wavelength. The goal of Doppler tomography is to infer 2D velocity distributions of the emitters from spectroscopic measurements. In fusion plasmas such spectra are measured by fast-ion D$_\alpha$ (FIDA) spectroscopy \cite{Heidbrink2004, Heidbrink2010, Geiger2011, Geiger2013, Geiger2014, Geiger, Salewski2014, Salewski} where fast ions are neutralized to become excited neutrals emitting $D_\alpha$-photons. The Doppler shift is determined by the phase angle of the fast ion at the time of the charge-exchange reaction.

The available measurements in astrophysical and fusion plasma Doppler tomography lead to different flavours of Doppler tomography. Since the two stars in an interacting binary orbit each other, we can view the binary at any phase angle in their orbit. The observations are made mostly using ground-based telescopes but sometimes also satellite-based telescopes such as the Hubble space telescope. Provided observations are obtained across a substantial fraction of the period of the binary system, spectra for various phase angles can be adequately sampled using time-series observations. These spectra change with the phase since several prominent features in the accretion disk are phase-locked to the binary. Astrophysical Doppler images are then inferred for two velocity coordinates in the orbital plane of the binary. The out-of-plane velocity component of matter in the disc is negligible as the disc is thin compared to its diameter. On the contrary, line radiation from fusion plasmas comes from many emitters at all phases, and hence spectra are not phase-resolved but phase-averaged. But since the ion velocity distribution function is to a good approximation rotationally symmetric about the strong and slowly varying local magnetic field, resolution of different phases is not necessary. Fusion plasma Doppler tomograms are imaged in velocity components parallel and perpendicular to the magnetic field. Astrophysical Doppler tomograms are 2D by assuming zero out-of-plane velocity whereas fusion plasma Doppler tomograms are 2D by assuming rotational symmetry.

\begin{figure}[tbp]
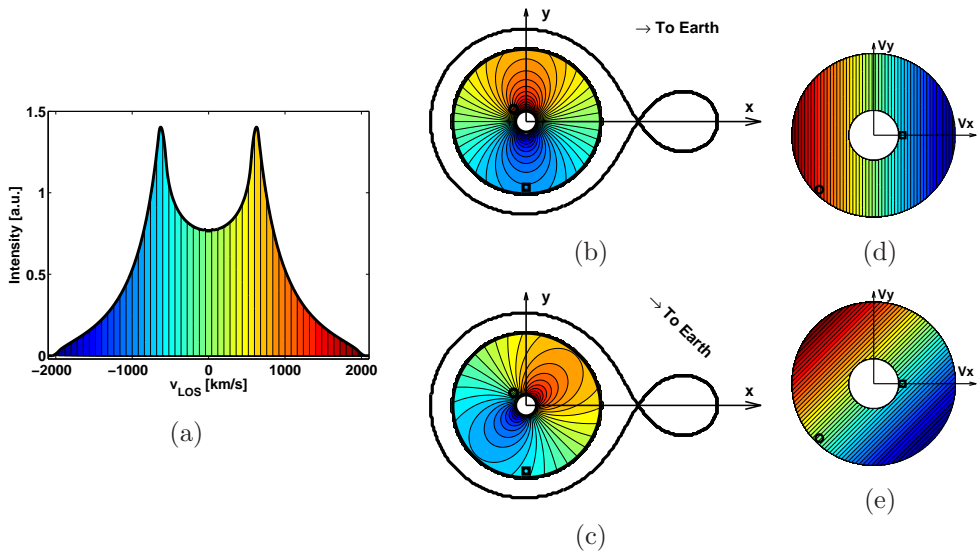

\centering
\begin{minipage}{0.35\linewidth}
\subfigure[]{
	\includegraphics[width=50mm]{figures/vlosspectrumdisc.epsc}
	}
\end{minipage}
\begin{minipage}{0.3\linewidth}
\subfigure[]{
	\includegraphics[width=45mm]{figures/position0.epsc}
	}
\subfigure[]{
	\includegraphics[width=45mm]{figures/position45.epsc}
	}	
\end{minipage}
\begin{minipage}{0.3\linewidth}
\subfigure[]{
	\includegraphics[width=25mm]{figures/velocity0.epsc}
	}
\subfigure[]{
	\includegraphics[width=25mm]{figures/velocity45.epsc}
	}
\end{minipage}

\caption{(a) Model spectrum from an accretion disc in LOS velocity space. Negative $v_{LOS}$ corresponds to blue-shift and positive $v_{LOS}$ to red-shift. (b)-(d) show where these $v_{LOS}$ appear on the disc in position space and in velocity space in the same colours as in (a). (b) and (c) show the disc and the Roche lobe of a binary in position space at different phases. The Roche lobe is an equipotential surface of the rotating two-body system. Matter within the Roche lobe of a star is gravitationally bound to this star. (d) and (e) show the same accretion disc in velocity space at the phases of (b) and (c), respectively. The circles and the squares illustrate the mapping from position space to velocity space.} \label{fig:VspacePspace}
\end{figure}

The incentives for imaging in velocity space are different in the two fields. In astrophysical Doppler tomography one is actually interested in the spatial structure of accretion discs. Each point of an accretion disc can be mapped onto velocity space by for example assuming flow velocities obeying Kepler's law which may, however, be a crude assumption. Such Keplerian mapping is illustrated in figure~\ref{fig:VspacePspace}. Here we observe that typical rotational speeds in the disc are much larger than the thermal speed of the atoms, so that the line broadening is mostly caused by the rotation of the accretion disc. Tomographic reconstructions in velocity space are preferable as they do not require any mapping assumptions and make the method applicable to a wide range of flow geometries including emission sources not originating within discs. In fusion plasma Doppler tomography, on the contrary, knowledge of the fast-ion phase-space distribution function $f(\textbf{u}, \textbf{x})$ itself is essential for the successful operation of a fusion power plant.


In this paper we discuss Doppler tomography in astrophysical and nuclear fusion applications. In section~\ref{sec:principles} we discuss general principles appearing in both applications as well as their differences. We compare the equations describing the projection onto the LOS, the forward models, and the most wide-spread inversion methods. In section~\ref{sec:astrophyiscs} we highlight achievements of astrophysical Doppler tomography, and in section~\ref{sec:fusion} we present fusion plasma Doppler images. In section~\ref{sec:discussion} we discuss the two applications of the Doppler tomography method in light of each other and what can be learned by comparison. In section~\ref{sec:conclusions} we draw conclusions.

\section{Principles of velocity-space tomography}
\label{sec:principles}
\subsection{Line-of-sight velocity}
Here we derive how $v_{LOS}$ relates to the astrophysical 2D velocity space of the orbital plane ($v_x, v_y$) and to the fusion plasma 2D velocity space ($u_\|, u_\perp$). Consider a coordinate system with unit vectors ($\hat{\textbf{u}}_x, \hat{\textbf{u}}_y, \hat{\textbf{u}}_z$) and components ($u_x, u_y, u_z$) in velocity space as illustrated in figure~\ref{fig:VspacePspace} where the $u_z$-axis is aligned with the rotation axis. The orientation of $u_x$ and $u_y$-axes is arbitrary for fusion plasma due to rotational symmetry. Let the LOS be at an inclination angle $i$ to the rotation axis and have an azimuthal angle $\phi$ from the $(\hat{\textbf{u}}_x, \hat{\textbf{u}}_z)$-plane, and let  $\gamma$ be a systemic or drift velocity along the LOS. Then the unit vector along the LOS $\hat{\textbf{v}}$ and the emitter velocity \textbf{u} are
\begin{eqnarray}
\label{eq:los}
\hat{\textbf{v}}&=& - \cos \phi \sin i \hat{\textbf{u}}_x + \sin \phi \sin i \hat{\textbf{u}}_y - \cos i \hat{\textbf{u}}_z,  \\
\textbf{u} &=& \gamma \hat{\textbf{v}} + u_x \hat{\textbf{u}}_x + u_y \hat{\textbf{u}}_y + u_z \hat{\textbf{u}}_z.
\end{eqnarray}
The projected velocity $v_{LOS}$ along the LOS from 3D velocity space is then \cite{Marsh1988}
\begin{eqnarray}
v_{LOS} = \gamma - u_x \sin i \cos \phi + u_y \sin i \sin \phi - u_z \cos i.
\label{eq:projectionfrom3D}
\end{eqnarray}
In the following we make further assumptions to deduce simplified projection equations for astrophysical and fusion plasma Doppler tomography. In astrophysics the inclination $i$ of the accretion disc is often unknown, and so one substitutes \cite{Marsh1988}
\begin{eqnarray}
\label{eq:transformUtoV}
v_x=u_x \sin i, \hspace{0.3cm} v_y=u_y \sin i, \hspace{0.3cm} v_z=u_z \cos i 
\end{eqnarray}
to get a projection equation not containing the inclination:
\begin{eqnarray}
\label{eq:projectionfromVxVyVz}
v_{LOS}=\gamma - v_x \cos \phi + v_y \sin \phi -v_z.
\end{eqnarray}
Further, the out-of-plane flow $v_z$ is assumed to be zero, and we arrive at the projection equation used in many astrophysical applications \cite{Marsh1988}:
\begin{eqnarray}
\label{eq:projectionfromVxVy}
v_{LOS}=\gamma - v_x \cos \phi + v_y \sin \phi.
\end{eqnarray}
In fusion plasma Doppler tomography, the magnetic field and the LOS vectors and hence the inclination $i$ are known, making transformation to $(v_x, v_y, v_z)$-coordinates unnecessary. To exploit rotational symmetry, we transform to cylindrical coordinates: 
\begin{eqnarray}
u_x=u_\perp \cos \bar{\phi}, \hspace{0.3cm} u_y=u_\perp \sin \bar{\phi}, \hspace{0.3cm} u_z=u_\|
\end{eqnarray}
so that the projection equation becomes
\begin{eqnarray}
v_{LOS} &=& \gamma - u_\perp \cos \phi \sin i \cos \bar{\phi} + u_\perp \sin \phi \sin i \sin \bar{\phi} - u_\| \cos i \nonumber \\
&=& \gamma - u_\perp \sin i \cos (\phi + \bar{\phi}) - u_\| \cos i. 
\end{eqnarray}
As the distribution is rotationally symmetric, we can choose $\phi=0$. Further, it is assumed that there is no systemic or drift velocity $\gamma$. Hence we obtain the usual projection equation used for Doppler tomography in fusion plasma
\begin{eqnarray}
v_{LOS} = - ( u_\| \cos i  + u_\perp \sin i \cos \bar{\phi}) \label{eq:projectionfromUpaUpe}
\end{eqnarray}
where normally the vector along the LOS is defined in the opposite direction so that the minus disappears \cite{Salewski2011, Salewski}. Thus the two projection equations are consistently derived but describe projections from different image planes and rely on different assumptions.

\subsection{Forward models}
The projection of an arbitrary 3D function $f_u^{3D}$ onto the LOS is \cite{Marsh1988, Salewski2011}
\begin{eqnarray}
f_{v,LOS}(v,\phi)&=& \int_{-\infty}^\infty \int_{-\infty}^\infty \int_{-\infty}^\infty f_u^{3D}(u_x, u_y, u_z) \delta \Bigg(v-v_{LOS} \Bigg)du_x du_y du_z .
\label{eq:projectionf3D}
\end{eqnarray}
We now reduce equation~\ref{eq:projectionf3D} by making the same assumptions as in the previous section. In astrophysics we transform the velocity coordinates to $(v_x, v_y, v_z)$ using equation~\ref{eq:transformUtoV} and substitute for $v_{LOS}$ using equation~\ref{eq:projectionfromVxVyVz}:
\begin{eqnarray}
f_{v,LOS}(v,\phi)&=& \int_{-\infty}^\infty \int_{-\infty}^\infty \int_{-\infty}^\infty f_v^{3D}(v_x, v_y, v_z) \nonumber \\ &\times & \delta \Bigg(v- \gamma + v_x \cos \phi - v_y \sin \phi + v_z \Bigg)dv_x dv_y dv_z .
\end{eqnarray}
Assuming the out-of-plane velocity to be negligible, we write 
\begin{eqnarray}
f_v^{3D}(v_x, v_y, v_z)=f_v^{2D}(v_x, v_y) \delta(v_z)  
\end{eqnarray}
and integrate over $v_z$ to find the common astrophysical 2D projection equation \cite{Marsh2001}
\begin{eqnarray}
\hspace{-0.8cm} f_{v,LOS}(v,\phi)&=& \int_{-\infty}^\infty \int_{-\infty}^\infty f_v^{2D}(v_x, v_y) \delta \Bigg(v- \gamma + v_x \cos \phi - v_y \sin \phi \Bigg)dv_x dv_y.
\end{eqnarray}
In fusion plasma physics, $u_z$ is allowed to be arbitrary, but we assume the $f_u^{3D}(u_x, u_y, u_z)$ to be rotationally symmetric so that it can be described by two coordinates ($u_\|, u_\perp$). We define a 2D velocity distribution function
\begin{eqnarray}
f_u^{2D}(u_\|, u_\perp)=\int_0^{2\pi
} f_u^{3D}(u_\|, u_\perp) u_\perp d\phi=2 \pi u_\perp f_u^{3D}(u_\|, u_\perp),
\label{eq:deff2D}
\end{eqnarray}
transform equation~\ref{eq:projectionf3D} to cylindrical coordinates and substitute equation~\ref{eq:deff2D} \cite{Salewski2011}
\begin{eqnarray}
\hspace{-2.5cm}f_{v,LOS}(v, i)&=& \int_{-\infty}^\infty \int_0^\infty  \frac{1}{2\pi} \int_0^{2 \pi} \delta (u_\| \cos i + u_\perp \sin i \cos \phi - v) d\phi  f_u^{2D}(u_\|, u_\perp) du_\perp du_\|. \nonumber \\
\end{eqnarray}
As one actually measures the number of photons in a small wavelength range rather than at one wavelength, the measurable quantity is the integral of $f_v$ over a small velocity range \cite{Salewski}. Noting that $f_u^{2D}(u_\| , u_\perp)$ does not depend on $v$, we arrive at the forward model used in fusion plasma Doppler tomography:
\begin{eqnarray}
f_{LOS}(v_1, v_2, i)&=&\int_{v_1}^{v_2} f_{v,LOS}(v, i) dv \nonumber \\
&=& \int_0^\infty \int_{-\infty}^\infty w (v_1, v_2, i, u_\|, u_\perp) f_u^{2D}(u_\| , u_\perp) d u_\|  d u_\perp
\label{eq:wfcDef}
\end{eqnarray}
where we have introduced a weight function $w$ in analogy with position-space tomography. The weight function can be explicitly calculated \cite{Salewski2011, Salewski}:
\begin{eqnarray}
w(v_1, v_2, i, u_\|, u_\perp) &=&\int_{v_1}^{v_2} \frac{1}{2\pi} \int_0^{2 \pi} \delta (u_\| \cos i + u_\perp \sin i \cos \phi - v) d\phi   dv \nonumber \\
&=&\int_{v_1}^{v_2} \frac{1}{\pi  u_\perp \sin i \sqrt{1-\Big(\frac{v-u_\| \cos i}{u_\perp \sin i}\Big)^2}} dv \nonumber \\
&=& \frac{1}{\pi}\Big(\arccos \frac{v_1-u_\| \cos i}{u_\perp \sin i} - \arccos \frac{v_2-u_\| \cos i}{u_\perp \sin i} \Big).
\label{eq:wfcDopplerUniform}
\end{eqnarray}
Weight functions in this form enable us construct a transfer matrix $W$ with which we can rapidly calculate the implied function $f_{LOS}$ from an arbitrary image in 2D velocity space $f_u^{2D}$. The forward model can be written as a matrix equation of the form
\begin{eqnarray}
\label{eq:transfermatrix}
F_{LOS}=W F_u^{2D}
\end{eqnarray}
where $F_{LOS}$ is a vector holding the measurements and $F_{u}^{2D}$ is a vector holding the image pixel values \cite{Salewski2011, Salewski2012, Salewski2013, Salewski2014}. Refined forward models accounting for Stark splitting, charge-exchange reaction probabilities and electron transition probabilities are discussed in reference \cite{Salewski}. In this paper we focus on the Doppler shift to emphasize the analogy between astrophysical and fusion plasma Doppler tomography.

\subsection{A rotationally symmetric accretion disc with no out-of-plane flow}
\begin{figure}
\centering
\includegraphics[width=70mm]{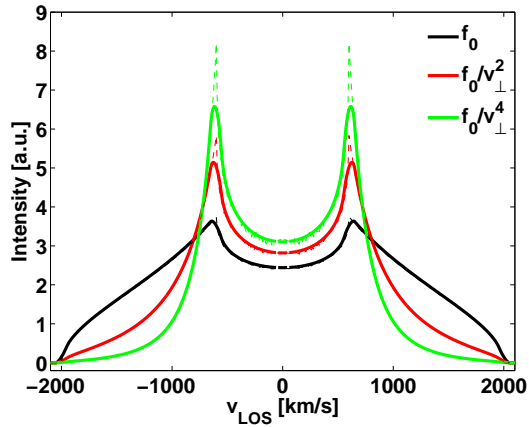}
\caption{Theoretical spectra of rotationally symmetric accretion discs with no out-of-plane flow and intensities following power laws of the form $f\sim f_0 / v_\perp^a$. Solid lines show the analytic formulas and dashed lines the numeric integration of the velocity image not accounting for finite resolution.}\label{fig:symmetricdiscs}
\end{figure}

Here we derive explicit formulas for the observable spectrum of a rotationally symmetric accretion disc with velocities from $v_{\perp 1}$ to $v_{\perp 2}$ and no out-of-plane flow so that $f_v^{2D}(v_\perp , v_\|)=f_v^{1D}(v_\perp)\delta(v_\|)$. Exploiting the analogy with fusion plasma Doppler tomography, we integrate equation~\ref{eq:wfcDef} over $v_\|$ using equation~\ref{eq:wfcDopplerUniform}:
\begin{eqnarray}
\label{eq:accreationdisc}
f_{LOS}(v)
&=&\int_0^\infty \frac{1}{\pi}\Big(\arccos \frac{v_1}{v_\perp} - \arccos \frac{v_2}{v_\perp} \Big) f_v^{1D}(v_\perp) dv_\perp.
\end{eqnarray}
We can evaluate the integral over $v_\perp$ by assuming a functional form of $f_v^{1D}$. Similar models used position coordinates \cite{Smak1969, Smak1981, Horne;1986, Horne1995} whereas we use velocity coordinates. In these earlier treatments power laws were assumed and then matched to experimental data.  As for Keplerian flow power laws in position space map to power laws in velocity space, we also take the emitted intensity to follow a power law of the form $f_v^{1D}(v_\perp)=f_0 / v_\perp^a$ between $v_{\perp 1}$ and $v_{\perp 2}$. Hence we find theoretical spectra for rotationally symmetric discs with no out-of-plane flow for $a=(0,2,4)$:
\begin{eqnarray}
\hspace{-2.3cm} f_0 \rightarrow f_{LOS}(v_1,v_2)&=& \frac{f_0}{\pi}\Bigg(v \hspace{1mm} \textrm{arctanh} \frac{1}{\sqrt{1-\Big( \frac{v}{v_\perp}\Big)^2}} - v_\perp \arccos \frac{v}{v_\perp} \Bigg) \Big\vert_{v_\perp=v_{\perp 1}}^{v_\perp=v_{\perp 2}} \Big\vert_{v=v_1}^{v=v_2}\\
\hspace{-2.3cm} \frac{f_0}{v_\perp^2} \rightarrow f_{LOS}(v_1,v_2)&=&  \frac{f_0}{\pi} \Bigg( \frac{1}{v_\perp} \arccos \frac{v}{v_\perp} - \frac{1}{v} \sqrt{1-\Big( \frac{v}{v_\perp}\Big)^2} \Bigg) \Big\vert_{v_\perp=v_{\perp 1}}^{v_\perp=v_{\perp 2}} \Big\vert_{v=v_1}^{v=v_2}\\
\hspace{-2.3cm} \frac{f_0}{v_\perp^4} \rightarrow f_{LOS}(v_1,v_2)&=&  \frac{f_0}{\pi} \Bigg( \frac{1}{3 v_\perp^3} \arccos \frac{v}{v_\perp} - \frac{1}{9v}\Big(\frac{1}{v_\perp^2}-\frac{2}{v^2} \Big) \sqrt{1-\Big( \frac{v}{v_\perp}\Big)^2} \Bigg) \Big\vert_{v_\perp=v_{\perp 1}}^{v_\perp=v_{\perp 2}} \Big\vert_{v=v_1}^{v=v_2}
\end{eqnarray}
These spectra are shown in figure~\ref{fig:symmetricdiscs}. The value of $a$ is unknown and could be found by matching experimental data. For $a=0$ the 1D velocity is uniform corresponding to a $1/v$ curve in 2D velocity space. For $a=4$ very little emission comes from the rapidly rotating regions of the accretion disc which cover only a small area close to the accretor in position space. The models reproduce the characteristic often observed double-peak. This illustrates the analogy between astrophysical and fusion plasma Doppler tomography as we derived the astrophysical observation from the fusion plasma formula. Further, the model gives direct insight in the relation between 2D velocity space of the accretion disc and the line-of-sight velocity space of the measurement, and it can be used to validate inversion algorithms.  The detailed shape of the double-peak is also influenced by the optical depth \cite{Horne;1986} and magnetohydrodynamic turbulence \cite{Horne1995}.

\subsection{Inversion methods}
\label{sec:algorithms}
Several inversion methods have been applied to solve the velocity-space tomography problem. Astrophysical applications usually apply the maximum entropy method \cite{Marsh1988} or filtered back-projection method \cite{Marsh2001}. Nuclear fusion applications of velocity space tomography have relied on the singular value decomposition \cite{Salewski2012, Salewski2013, Salewski2014}, an iterative technique akin to the back-projection method \cite{Salewski2011}, and a maximum entropy method \cite{Stagner2013}. In all methods velocity space is divided into many elements or pixels. In the maximum entropy method we calculate synthetic data for possible images and quantify the difference between the synthetic data with the measured data by a goodness-of-fit parameter $\chi^2$.  $\chi^2$ could be decreased to very low values by changing the image, but this tends to amplify noise in the images. Instead we set a target $\chi^2$ such that we judge the synthetic and the measured data to be consistent. The reduced $\chi^2$ should then be of order one but the precise value is open for discussion. Out of the many tomograms that achieve this target $\chi^2$ one selects the one with maximum entropy which can be found using Lagrange multipliers in an iterative procedure  \cite{Skilling1984}. The standard definition of entropy
\begin{eqnarray}
S=-\sum p_i \ln p_i, \hspace{0.3cm} p_i = f_i / \sum_j f_j
\end{eqnarray}
selects the most uniform image \cite{Marsh1988}. In astrophysical Doppler tomography one frequently instead uses a modified entropy
\begin{eqnarray}
S=-\sum p_i \ln \frac{p_i}{q_i},  \hspace{0.3cm} p_i = f_i / \sum_j f_j,  \hspace{0.3cm} q_i = D_i / \sum_j D_j
\end{eqnarray}
where D is a default image \cite{Marsh1988}. Prior information can be encoded in the default image which can be set to be for example axisymmetric or a heavily blurred version of the image. Such adaptive defaults are better than axisymmetric defaults as the Doppler images have well-defined spots of emission and discs can be strongly asymmetric. An advantage of the maximum entropy method is the enforced positivity that reduces high frequency jitter in the image which is often found in linear methods. 

The filtered back-projection method is a linear method in which the inversion is computed in two steps. First the spectra are filtered to damp high frequency components which would otherwise lead to jitter in the tomogram. This step also blurs the tomogram. The filtered profiles $\tilde{f}(v, \phi)$ are
\begin{eqnarray}
\tilde{f}(v, \phi)=\int F(\hat{v},v)f(v-\hat{v},\phi)d\hat{v}.
\end{eqnarray}
where $F$ is the filter function. The second step is the so-called back-projection which is
\begin{eqnarray}
f(v_x, v_y)=\int_0^{0.5} \tilde{f}(\gamma -v_x \cos 2 \pi \phi' + v_y \sin 2 \pi \phi', \phi')d\phi'.
\end{eqnarray} 
In this method each image value is found by integrating the 2D function $\tilde{f}(v, \phi)$ over the sinusoidal path which would be traced out by a bright light source with the velocity coordinates of the image point. One may also regard this as smearing the filtered profile measured at the angle $\phi$ across the image at the same angle $\phi$.

The singular value decomposition method is another linear method, in which we formulate a forward model based on weight functions as a matrix equation. The tomographic inversion is then given by
\begin{equation}
F^+=\hat{W}^+ \hat{F}_{LOS}.
\end{equation}
$\hat{W}^+$ is the truncated Moore-Penrose pseudoinverse found by singular value decomposition of the transfer matrix $\hat{W}$ from equation~\ref{eq:transfermatrix}.

\section{Doppler tomography of accreting binary stars}
\label{sec:astrophyiscs}
\begin{figure}[htbp]
\centering
	\subfigure[Spectra]{%
	\includegraphics[width=50mm]{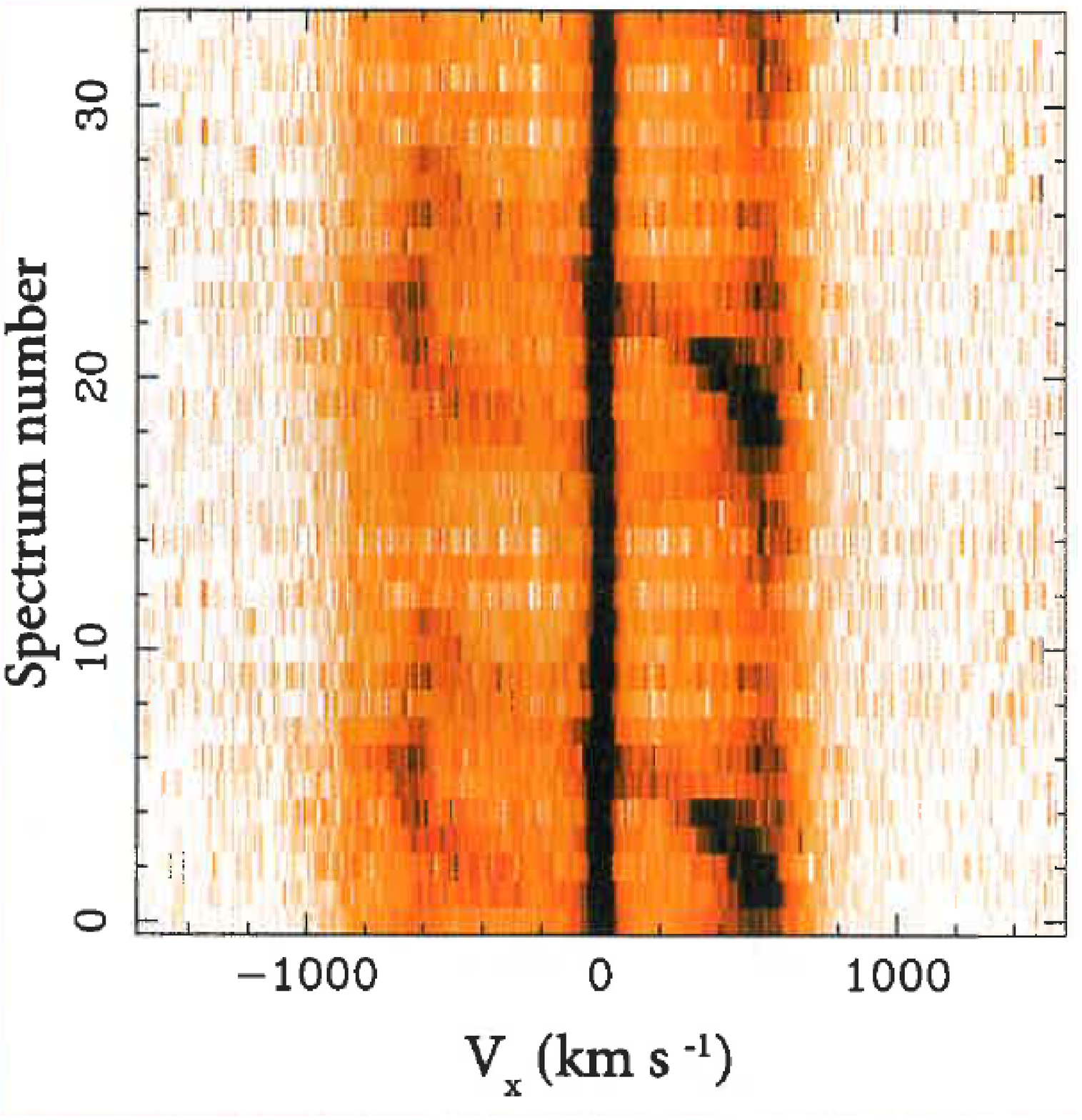}
	}
	\subfigure[Doppler image]{%
	\includegraphics[width=50mm]{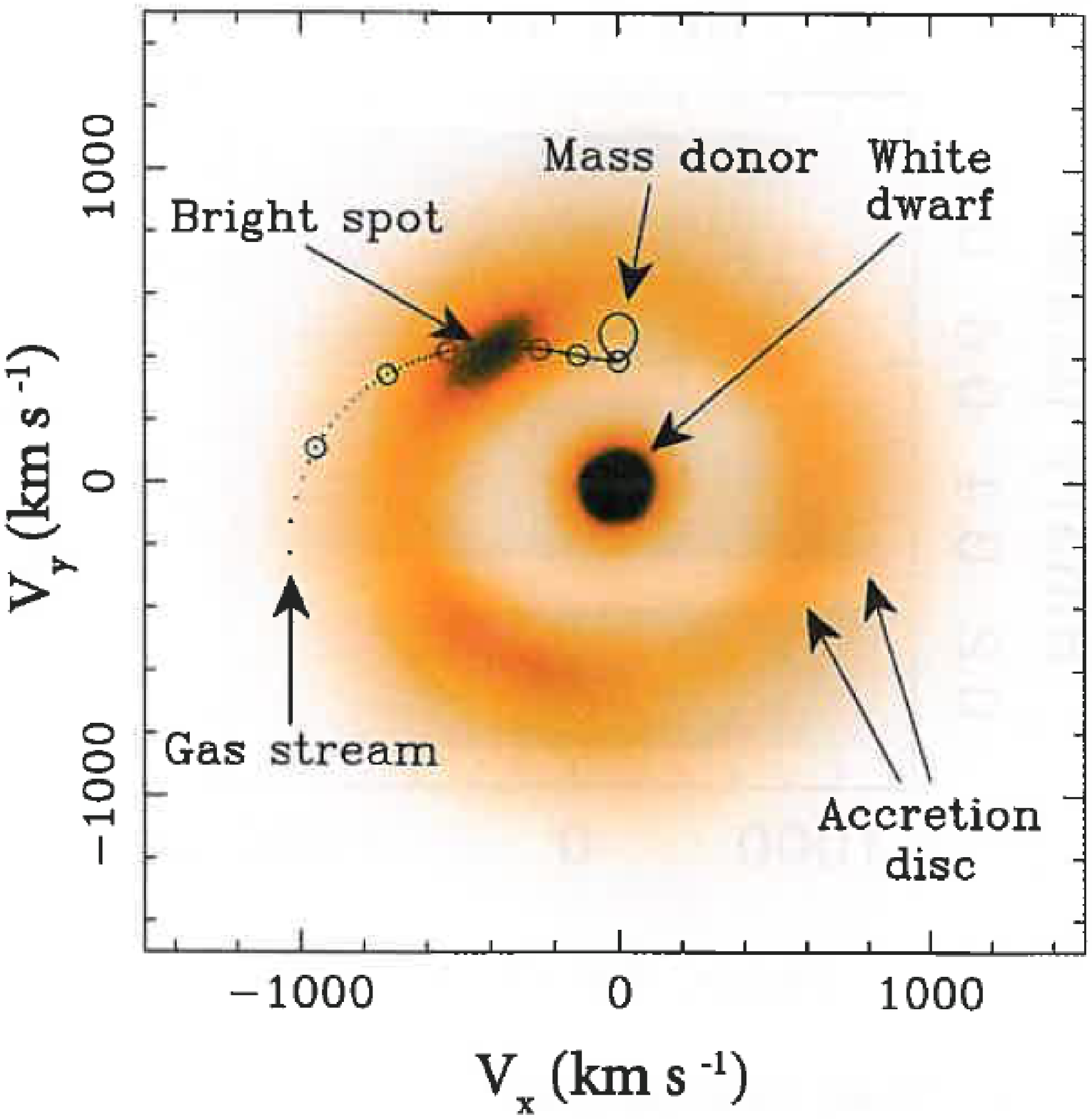}
	}
\caption{(a) The observed line emission from CE315 as a function of the binary orientation and projected velocity which is here called $V_x$ \cite{Marsh2005}. Each of the 34 spectra is measured for a new phase angle of the binary, corresponding to a new line-of-sight. (b) The corresponding Doppler image reveals a bright spot where the gas stream from the donor to the accretor hits the accretion disc. The dotted line shows a calculated ballistic trajectory of the gas stream where the circles show 10\% steps of the distance which the gas stream covers \cite{Marsh2005}. The $V_x$-coordinate in (a) corresponds to the $V_x$-coordinate in (b) for one particular phase.} \label{fig:CE315}
\end{figure}

Astrophysical Doppler tomography is a standard technique to image accretions discs of binary star systems such as cataclysmic variables \cite{Morales-Rueda2004}, Algols \cite{Richards2004}, and X-ray binaries \cite{VrtiLek2004} including neutron stars \cite{Steeghs2002, Bassa2009} and black holes \cite{Marsh1994, Neilsen2008}. It is also very useful to map the magnetically controlled accretion stream in polars where the strong magnetic fields prevent formation of an accretion disc \cite{Schwope2004} or in intermediate polars \cite{Bloemen2010}. Often spectra of the strong emission lines from $H$ and $He$ are measured, but recently spectra of the Ca\textsl{II} line even revealed the presence of the faint donor star \cite{VanSpaandonk2010}. Here we highlight two particularly instructive achievements of astrophysical Doppler tomography. Figure~\ref{fig:CE315} shows the observed time-series of emission line profiles, so-called trailed spectra, and the corresponding Doppler tomogram of the interacting binary CE315. The maximum entropy method was used for the inversion of the 34 measured spectra. The strong emission near 0 km/s comes from the compact, massive white dwarf. The trailed spectra in figure~\ref{fig:CE315}a  have a half-width of about 1000 km/s. They show the characteristic S-curve of a bright source of emission that is phase-locked to the binary. Figure~\ref{fig:CE315}b shows the Doppler image constructed from the trailed spectra. The accretion disk appears as a ring with velocities between 400 km/s and 1000 km/s. The highest speeds show emission from the inner edge of the disc in position space that is close to the white dwarf. The lowest speeds show emission from the outer edge of the disc in position space. The accreting white dwarf sits at the center of the Doppler image. The mass donor sits at $v_y=400$~km/s and by definition of the coordinate system at $v_x=0$. The Doppler image reveals a bright spot causing the S-curve in the trailed spectra. At this location the gas flow from the donor star to the white dwarf hits the disc.  In the Doppler image this gas flows from the L1 Lagrange point between the donor and the accretor to the high-speed outer edge of the disc which, in position space, corresponds to the inner edge of the disc close to the accretor. 

\begin{figure}[htbp]
\centering
	\includegraphics[width=80mm]{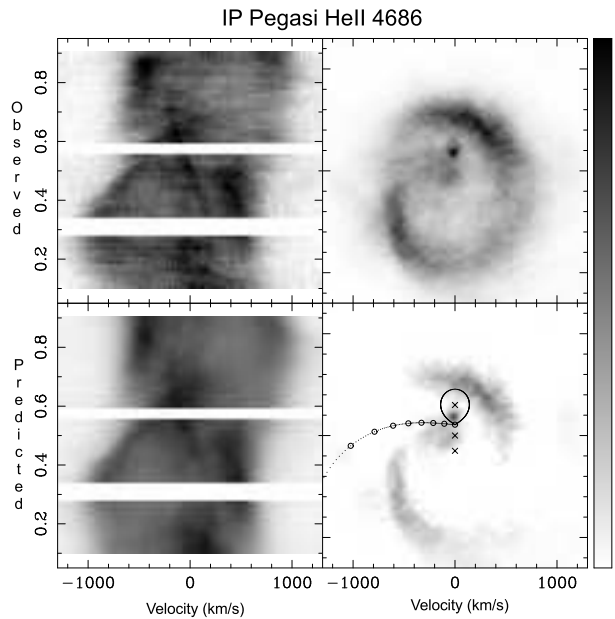}
\caption{In the upper panels we show the observed line emission from IP Peg as a function of the binary orientation and projected velocity (left) and corresponding Doppler image revealing spiral arms in the accretion disc (right) \cite{Harlaftis1999}. The binary phase in the left panel runs from 0 to 1 during a period of the binary. Both axes of the Doppler image in the right panel are in identical units. The velocities are defined analogous to those in figure~\ref{fig:CE315}. The lower panels show a numerical simulation of the disc (right) and the implied observable line emission (left). The gas stream is marked as in figure~\ref{fig:CE315}.} \label{fig:IPPeg}
\end{figure}

In figure~\ref{fig:IPPeg} we show an example of the very variable spectra observed over the binary phase of the binary IP Peg \cite{Steeghs1997, Steeghs1999, Harlaftis1999}. The Doppler images reveal that these variable spectra, which would otherwise be very hard if not impossible to interpret, are due to spiral arms in the accretion disc \cite{Steeghs1997}. A numerical simulation and derived synthetic observations match respectively the Doppler image and observations well. These spiral arms are phase-locked to the binary suggesting they are some form of tidal wave \cite{Steeghs1997}, perhaps tidally induced shock waves \cite{Harlaftis1999}. However, the nature of the spiral arms is still controversial. Spiral arms have been observed in many other binaries, e.g. \cite{Bloemen2010}. Other asymmetric features in the discs have also been observed, e.g. eccentricity \cite{Roelofs2006} or alternating radial flow velocities \cite{Copperwheat2012}. These examples illustrate how the complex emissivity profiles observed from binaries can be conveniently mapped into images that are much more straightforward to interpret and at the same time offer quantitative tests against models.

\section{Doppler tomography of fast-ion velocity distribution functions}
\label{sec:fusion}
Tomography in position space is a standard analysis method in nuclear fusion research \cite{Ertl1996, Anton1996} just as in other fields throughout physical and medical sciences \cite{A.C.KakandM.Slaney1988, G.T.Herman2009}. Fusion plasma Doppler tomography has been studied theoretically for some years \cite{Egedal2004, Salewski2011, Salewski2012, Salewski2013}. The method has been theoretically demonstrated and made tractable by formulating the forward model in terms of weight functions \cite{Salewski2011} previously used to estimate the velocity-space sensitivity of FIDA measurements \cite{Heidbrink2007, Heidbrink2010, Salewski}. 

\begin{figure}[htbp]
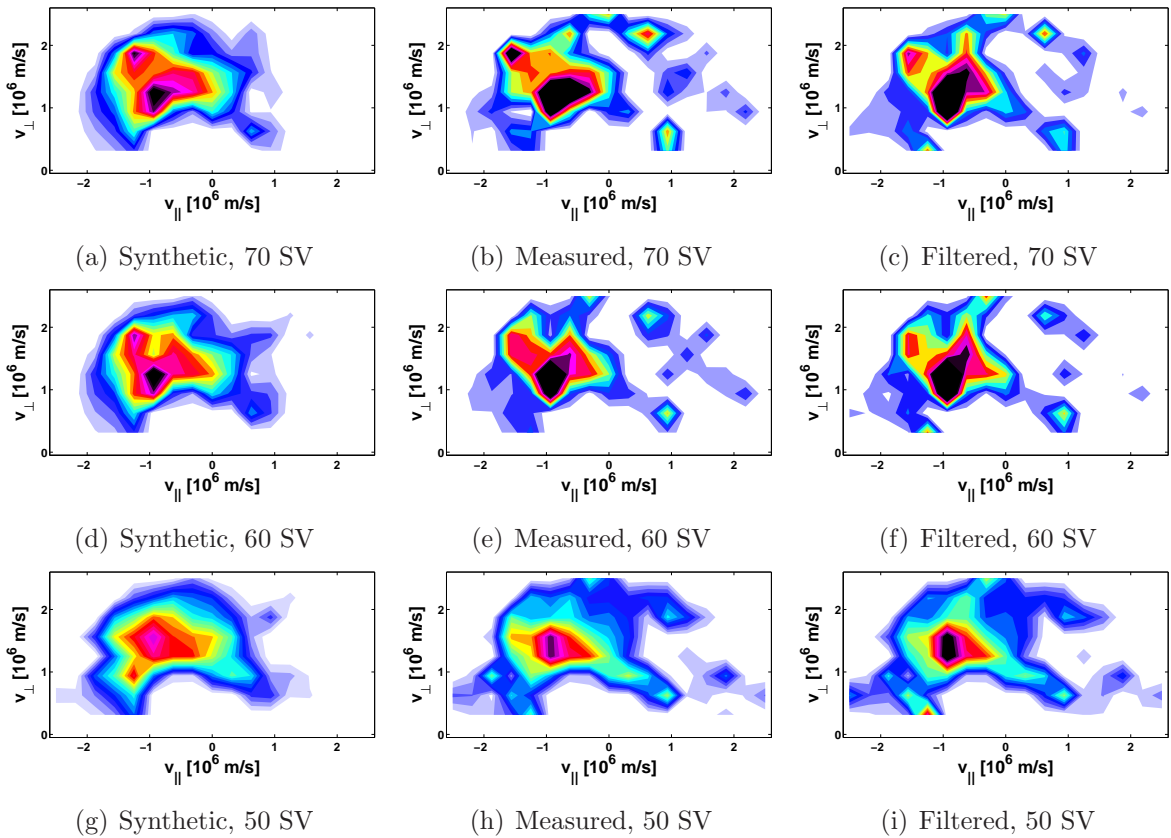

\centering
	 \subfigure[Synthetic, 70 SV]{%
	\includegraphics[width=50mm]{figures/reconstruction70.epsc}
	}
	\subfigure[Measured, 70 SV]{%
	\includegraphics[width=50mm]{figures/tomo70.epsc}
	}
	\subfigure[Filtered, 70 SV]{%
	\includegraphics[width=50mm]{figures/tomofiltered70.epsc}
	}
    \subfigure[Synthetic, 60 SV]{%
	\includegraphics[width=50mm]{figures/reconstruction60.epsc}
	}
	\subfigure[Measured, 60 SV]{%
	\includegraphics[width=50mm]{figures/tomo60.epsc}
	}
	\subfigure[Filtered, 60 SV]{%
	\includegraphics[width=50mm]{figures/tomofiltered60.epsc}
	}
    \subfigure[Synthetic, 50 SV]{%
	\includegraphics[width=50mm]{figures/reconstruction50.epsc}
	}
	\subfigure[Measured, 50 SV]{%
	\includegraphics[width=50mm]{figures/tomo50.epsc}
	}
	\subfigure[Filtered, 50 SV]{%
	\includegraphics[width=50mm]{figures/tomofiltered50.epsc}
	}
\caption{Doppler images from synthetic FIDA spectra (left column), from measured FIDA spectra (center column), and from filtered, measured FIDA spectra (right column). The number of singular values is 70 in the uppermost row, 60 in the center row, and 50 in the bottom row.} \label{fig:fvpavpetomo}
\end{figure}

Figure~\ref{fig:fvpavpetomo} shows Doppler images of the fast-ion velocity distribution function in ASDEX Upgrade discharge 29578 on $17 \times 8$ grid points. Here we study the number of singular values and the effect of filtering the spectra as in the filtered back-projection method. From the uppermost row to the bottom row the number of singular values in the tomogram decreases. In the left column we use synthetic measurements, in the center column actual measurements, and in the right column filtered actual measurements. 

We measured in three FIDA views and used the singular value decomposition method to invert the spectra \cite{Salewski2014}. We used experimentally accessible parts of the spectrum with red- and blue shifts with a wavelength resolution of 0.1~nm over 16~nm. Of the resulting $3 \times 160$ measurement points, 217 were not obscured by other features in the FIDA spectra and covered the velocity-space region of interest. The inclination angles $i$ of the three FIDA LOS are $12^\circ$, $69^\circ$ and $156^\circ$. The left column in figure~\ref{fig:fvpavpetomo} shows reconstructions from synthetic FIDA measurements computed from a simulated fast-ion velocity distribution function using the FIDASIM code \cite{Heidbrink2011}. Figure~\ref{fig:fvpavpetomo}a closely matches simulations which we show in reference~\cite{Salewski2014}. 

This distribution function is typical for fast ions generated by neutral beam injection, and its form can be explained by classical slowing down due to collisions. Neutral deuterium atoms at $E=60$~keV are injected and ionized to $D$ ions, forming a peak at about $(v_\|, v_\perp) \approx (-1,2)\times 10^6$~m/s. $D_2$ and $D_3$ in the neutral beam lead to further injection peaks at $E/2=30$~keV and $E/3=20$~keV which are merged to form the second, larger peak at about $(v_\|, v_\perp) \approx (-1,1)\times 10^6$~m/s. These injection peaks are the sources of fast ions which then slow down due to collisions. In collisions with electrons the ions lose energy while their pitch $p=-V_\|/\sqrt{V_\|^2+V_\perp^2}$ does not change significantly. In collisions with ions the pitch also changes.

The Doppler images of the synthetic measurements show as expected that the larger the number of singular values, the finer features of the functions can be reconstructed but the more the noise is amplified. 50 singular values are not enough to recover the two peaks in the functions whereas for 60-70 singular values the two peaks appear. For more than 70 singular values the Doppler images become hard to interpret due to noise. There is no obvious objective rationale how to choose the number of singular values, and this is a weakness of the method. A possible remedy could be the L-curve technique \cite{Hansen1992}. The center column in figure~\ref{fig:fvpavpetomo} shows Doppler images from actual FIDA measurements at ASDEX Upgrade. The large-scale shape of the function including the location of the beam injection peaks are reproduced well if 60-70 singular values are used as predicted. The right column shows Doppler images calculated from filtered spectra as in the filtered back-projection method. We use a box filter with a stencil of three points. The filter decreases the amplitude of the jitter as expected. 

\begin{figure}[htbp]
\centering
	\includegraphics[width=90mm]{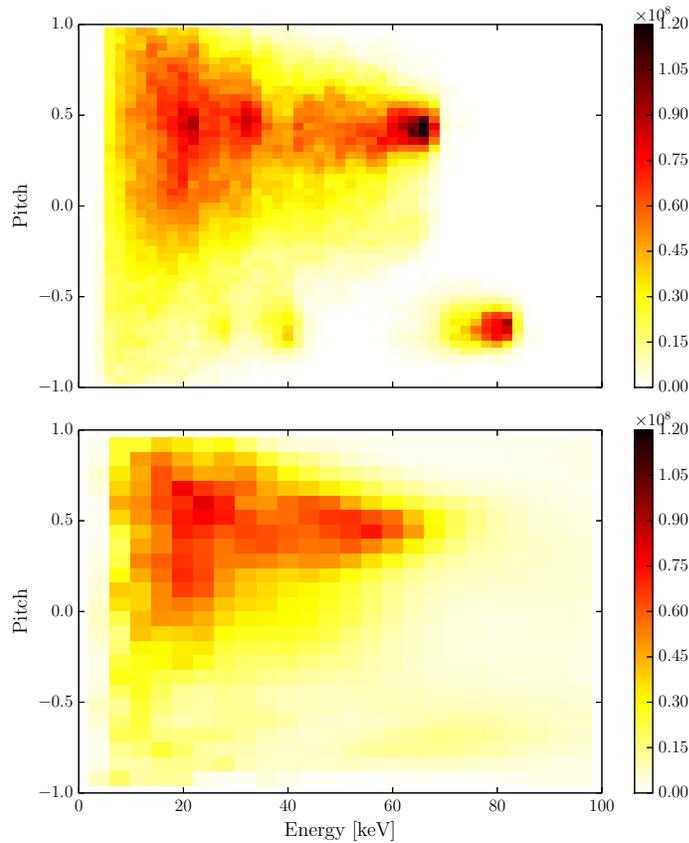}
\caption{Simulation (up) and Doppler image (below) from synthetic measurements of a fast-ion velocity distribution function at the tokamak DIII-D. Here the function is presented in the widespread (energy, pitch)-coordinates with $E=\frac{1}{2}  m v^2$ and $p=-v_\| / v$.} \label{fig:DIII-D}
\end{figure}

In figure~\ref{fig:DIII-D} we present a maximum entropy reconstruction of a fast-ion velocity distribution typical for the tokamak DIII-D on a $25 \times 25$ element energy-pitch grid from noisy synthetic measurements. Here we used four views with viewing angles of $8^\circ$, $19^\circ$, $45^\circ$ and $95^\circ$, a mean signal-to-noise ratio of 20:1, and a wavelength resolution of 0.14~nm over 14~nm, which corresponds to a total of $4 \times 100$ measurement points in the spectra. This simulated distribution assumes a 60 keV NBI in co-current direction (positive pitch) and an 80~keV NBI in counter-current direction (negative pitch) which generate the dominant beam injection peaks. Further peaks appear at half and third energies for both beams. The overall shape of the reconstruction matches the original function used to compute the synthetic measurements well, in particular for positive pitch. The peak at 80~keV at negative pitch in the reconstruction is barely visible, blurred and much smaller than in the original function. In the actual experiment the spectrum will be obscured by other emissions at small Doppler shifts. An experimental demonstration of the Doppler tomogaphy method on DIII-D with these or similar parameters is in preparation.

\section{Discussion}
\label{sec:discussion}
Doppler tomography in nuclear fusion research and astrophysics rely on the same techniques, but substantially different measurement data lead to different implementations of the method. The imaging plane is the orbital plane in astrophysics. The reduction from 3D is achieved by assuming that the out-of-plane velocity components are much smaller than the flow within the binary plane. Several researchers have attempted 3D imaging using all three velocity coordinates, though this is a much less well-constrained inversion problem from the observable time-series of 1D spectra. The assumption of rotational symmetry in fusion plasmas is rather good, so it would likely not lead to new insight to do 3D imaging in velocity space. Nevertheless, the inclusion of a spatial dimension in fusion plasma Doppler tomography would likely improve the inference. 

Astrophysical and fusion plasma Doppler tomography are photon-starved compared with many position-space tomography applications: The data is often scarce and the signal-to-noise ratio unfavourable. A spectrum in astrophysical Doppler tomography is analogous to a set of measurements with parallel or fanned beams along a LOS in position-space tomography. In astrophysical Doppler tomography the number of spectra or LOS's is limited by the signal-to-noise ratio, so that typically much fewer LOS's are used in astrophysical Doppler tomography (tens) than in position-space tomography (hundreds). In fusion plasmas actually only one LOS would be necessary for a measurement without noise since the fast-ion velocity distribution function is rotationally symmetric to a good approximation. However, due to noise in practise we need to use all available fast-ion measurements. So far three LOS's have been used, and this may be increased to seven or eight LOS's in the future. In fusion plasma Doppler tomography at ASDEX Upgrade, we can combine the FIDA measurements with other measurements \cite{Salewski2013} such as collective Thomson scattering \cite{Meo2008, Salewski2010a, Meo2010, Furtula2012, Salewski2011}, neutron emission spectroscopy or neutron yield measurements \cite{Giacomelli2011, Tardini2012, Tardini2013, Jacobsen2014, Jacobsen} or gamma-ray spectroscopy \cite{Nocente2012}. Similar combinations are possible at the tokamak DIII-D \cite{Heidbrink2007, Stagner2013}, the stellarator LHD \cite{Kubo2010, Nishiura2014, Ito2010} or the spherical tokamak MAST \cite{Michael2013a, Jones2013, Jones2014} as well as the next step fusion experiment ITER \cite{Salewski2008, Salewski2009a, Salewski2009, Kappatou2012, Bertalot2012, Chugunov2011}. In astrophysical Doppler tomography, one can use several emission lines from various elements such as hydrogen, helium, or calcium \cite{VanSpaandonk2010}.

The astrophysical Doppler tomography allows for a systemic velocity $\gamma$ along the LOS. This velocity is analogous to perpendicular drift in tokamak plasmas, such as the often dominant $\textbf{E} \times \textbf{B}$-drift or the often smaller grad-B, curvature or polarization drifts. Since parallel velocities are allowed in fusion plasma Doppler tomography, any drift velocities parallel to the magnetic field can already be handled, but not perpendicular drift velocities. It should be possible and beneficial to introduce a perpendicular drift velocity in the fusion plasma Doppler tomography approach as well, in particular when applied to the thermal ion population. This would possibly allow us to infer the perpendicular drift velocity and would probably also improve the Doppler image itself.

If there is a significant magnetic field, the line emission has finer structure. A moving D-atom in a magnetic field experiences an electric field in its own rest frame which causes the Balmer alpha line to split into 15 lines. This is referred to as Stark splitting. Stark shifts are usually substantially larger than Zeeman shifts which occur due to the magnetic field. Stark shifts are routinely accounted for in fusion plasma Doppler tomography by calculating the emission from the 15 lines and summing over them whereas Zeeman shifts are neglected. The effect of Stark splitting is significant for fusion plasmas \cite{Salewski}. In astrophysical Doppler tomography, Stark and Zeeman shifts have so far been neglected in Doppler tomography in binaries due to the large bulk velocities of the gas in the disc though Zeeman Doppler imaging has been successfully applied to resolve the stellar surfaces of magnetic stars. If the magnetic fields are strong, Stark and Zeeman shifts might also be a nuisance for Doppler tomography in binaries. They could be taken into a account in astrophysical Doppler tomography in binaries analogous to fusion plasma Doppler tomography. 

Finally, the inversion algorithms are readily transferable between astrophysical and nuclear fusion Doppler tomography. The formulation as a matrix problem is possible in astrophysics. Maximum entropy inversion algorithms are already used in both fields. The filtered back-projection method is widely used in astrophysics. A closely related method also using back-projection has been used in nuclear fusion Doppler tomography \cite{Salewski2011}. In this paper we borrowed the idea to filter the spectra and found that improvements may be possible with this technique. Filtering the measurement data is also a common technique in position-space tomography. Improvement of inversion algorithms will clearly benefit both fields.

\section{Conclusions}
\label{sec:conclusions} 
We outline basic principles of astrophysical and fusion plasma Doppler tomography by deriving projection equations and forward models from their common 3D framework. This enables us to derive the shape of observed spectra of light coming from accretion discs from the forward model of fusion plasma Doppler tomography. We present inversions of filtered measured spectra from the tokamak ASDEX Upgrade with the singular value decomposition method and of synthetic spectra with the maximum entropy method in preparation of Doppler tomography on the tokamak DIII-D. Prominent astrophysical Doppler images are discussed and compared with simulations. We already highlighted an example where an idea from one discipline was applied to the other. We further find that an inclusion of a perpendicular drift velocity in the fusion plasma forward model analogous to the systemic velocity of the binary in astrophysics will be valuable. Further, Stark and Zeemann splitting have so far been neglected in astrophysical Doppler tomography of binaries whereas Stark splitting is routinely accounted for in fusion plasma Doppler tomography. One could introduce similar models for the line splitting in astrophysical Doppler tomography even though additional models of the magnetic field would be required. Similar approaches have been successfully applied to image stellar surfaces of magnetic stars. Ideas in the inversion algorithms are readily transferable, for example the formulation as matrix equation used for fusion plasmas, filtering in linear methods to reduce noise, or different formulations of the entropy. In conclusion, using Doppler tomography we can conveniently map measured spectra into images that are much more straightforward to interpret and at the same time offer quantitative tests against models.

\section*{Acknowledgments}
This project has received funding from the Euratom research and training programme 2014-2018 and from the European Union’s Horizon 2020 research and innovation programme under grant agreement number 633053. The views and opinions expressed herein do not necessarily reflect those of the European Commission.

\section*{References}
\bibliography{salewski2014bibshort}



\end{document}